\documentclass[aps,prl,twocolumn,superscriptaddress,showpacs]{revtex4}

\usepackage{amsmath}
\usepackage{graphicx}

\newcommand{\be}{\begin{equation}}
\newcommand{\ee}{\end{equation}}
\newcommand{\mb}[1]{\mathbf{#1}}
\newcommand{\nn}{\nonumber}
\newcommand{\Journal}[4]{#1 \textbf{#2}, #3 (#4)}
\newcommand{\cd}{Cd$_2$Re$_2$O$_7$}

\begin{document}

\title{Goldstone-Mode Phonon Dynamics in the Pyrochlore Cd$_2$Re$_2$O$_7$}

\author{C. A. Kendziora$^{*}$}
\affiliation{Naval Research Laboratory, Washington, D.C. 20375-6365, USA}

\author{I. A. Sergienko$^{*}$}
\affiliation{Dept. of Physics and Physical Oceanography, 
Memorial University of Newfoundland, St. John's, NL, A1B 3X7, Canada}
\affiliation{Condensed Matter Sciences Division, Oak Ridge
National Lab., Oak Ridge, TN 37831, USA}

\author{R. Jin}
\affiliation{Condensed Matter Sciences Division, Oak Ridge
National Lab., Oak Ridge, TN 37831, USA}

\author{J. He}
\affiliation{Dept. of Physics and Astronomy, The University of Tennessee,
Knoxville, TN 37996, USA}

\author{V. Keppens}
\affiliation{Dept. of Materials Science and Engineering, The University of 
Tennessee, Knoxville, TN 37996, USA}

\author{B. C. Sales}
\affiliation{Condensed Matter Sciences Division, Oak Ridge
National Lab., Oak Ridge, TN 37831, USA}

\author{D. Mandrus}
\affiliation{Condensed Matter Sciences Division, Oak Ridge
National Lab., Oak Ridge, TN 37831, USA}

\begin{abstract} 
We have measured the polarized Raman scattering spectra of \cd, the first superconducting pyrochlore, as a function of temperature. For temperatures below the cubic-to-tetragonal structural phase transition (SPT) at 200K, a peak with B$_{1}$ symmetry develops at zero frequency with divergent intensity. We identify this peak as the first observation of the Goldstone phonon in a crystalline solid. The Goldstone phonon is a collective excitation that exists due to the breaking of the continuous symmetry with the SPT. Its emergence coincides with that of a Raman-active soft mode. The order parameter for both features derives from an unstable doubly-degenerate vibration (with E$_{u}$ symmetry) of the O1 atoms which drives the SPT.

\end{abstract}


\maketitle

The Goldstone theorem, originally formulated in particle physics, states
that there exists a massless particle if a continuous symmetry is spontaneously
broken~\cite{Goldstone62,Ryder96}. 
In the condensed state of matter, the massless particle corresponds to a 
collective excitation with wave vector $\mb k$ and frequency 
$\omega(\mb k\rightarrow 0) \rightarrow 0$. One simple example in which 
Goldstone modes are found is the Heisenberg model, the Hamiltonian of which
is invariant under simultaneous rotation of all spins in the lattice.
Indeed, experimentally Goldstone-like magnons have been found in a number 
of magnetic systems undergoing both finite 
temperature and quantum phase transitions~\cite{Ruegg04}. 
Goldstone vibrational modes are expected in ferroelectric liquid crystals due to the 
isotropy of the high-symmetry parent phase~\cite{Chandra92}. In contrast, 
such a behavior 
is not expected in crystalline solids because the nonlinear contributions to
the lattice vibrational Hamiltonian are usually highly anisotropic for 
degenerate phonons~\cite{Acoustic}.

\begin{figure}
\includegraphics[width = 3 in]{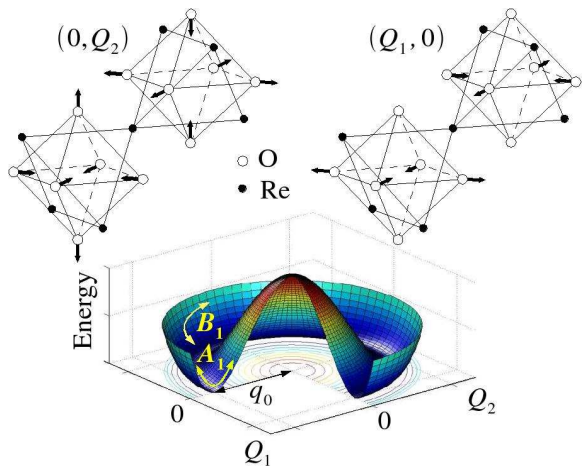}
\caption{\label{atoms}
(Color online)
(Top) Displacements of O1 oxygen atoms corresponding to the $E_u$ soft 
mode. In the cubic pyrochlore structure the O1 oxygen atoms form octahedra, 
the centers of which coincide with the centers of Re tetrahedra.
(Bottom) The effective potential of the soft mode in the broken-symmetry structure
($\alpha < 0$, $\beta>0$). Shown are the types of motion corresponding to the $A_1$ and 
$B_1$ modes.}
\end{figure}

In this Letter, we report polarized Raman investigations of the superconducting
oxide \cd~\cite{Hanawa01,Sakai01,Jin01}, which shows evidence of  
Goldstone phonon mode behavior. To the best of our knowledge, this is the 
first observation of a Goldstone \emph{optical phonon} mode in a crystalline solid. 
\cd, a cubic pyrochlore (space group $Fd\bar3m$) at room temperature, exhibits 
two structural phase transitions (SPT) at the temperatures of 200 K and 
120 K~\cite{Castellan02,Yamaura02,Arai02}. 
The corresponding low temperature space groups are $I\bar4m2$ and $I4_122$, both
tetragonal and non-centrosymmetric. 
The order parameter corresponds to a Brillouin zone centered ($\mb k=0$) doubly
degenerate phonon of $E_u$ symmetry dominated by the displacements of oxygen 
atoms occupying position 48(f) of $Fd\bar3m$ space group~\cite{Sergienko03,Sergienko04}. 
These atoms, which comprise 6 of the 7 oxygen per formula unit, are collectively referred 
to as the O1 atoms to distinguish them from the O2 atom at the 8(a) site. The displacements
of the O1 oxygen atoms corresponding to the soft mode are shown in Fig~\ref{atoms}. The Cd 
and O2 atoms are not shown.

The effective Hamiltonian for the soft mode is
\be
\label{Heff}
H_\text{eff}=\frac{P_1^2+P_2^2}{2 M} + \frac \alpha 2 (Q_1^2+Q_2^2) + 
\frac \beta 4 (Q_1^2+Q_2^2)^2, 
\ee
where $(Q_1,Q_2)$ are the symmetric coordinates, $(P_1,P_2)$ are the 
corresponding momenta and $M$ is the effective mass. The coefficients $\alpha$ 
and $\beta$ are in general temperature dependent due to the coupling to all the
other phonon modes of the crystal. A crucial feature of $H_\text{eff}$ is the absence of 
anisotropic terms in the third-order or fourth-order part of the potential energy. We note 
that this is not an assumption but an exact result of the symmetry 
analysis~\cite{Sergienko03}. The first anisotropic contribution is of the form 
$\frac \gamma 6 (Q_1^3-Q_1Q_2^2)^2$. This term is important in lifting the degeneracy 
between the tetragonal phases. However, it is not expected to be significant for the 
dynamical properties at least close to the second-order phase transition at 200 K. 
The smallness of $\gamma$ is supported by the fact that the observation of the 120 K SPT 
is sample dependent and is related to crystal imperfections~\cite{Lu05} as well as by
first-principles calculations~\cite{Sergienko04}.
Consequently, $H_\text{eff}$ is invariant with respect to arbitrary 
``rotations'' in the order parameter space $(Q_1,Q_2)$, as depicted in  
Fig.~\ref{atoms}. When the cubic symmetry is broken at 200 K, this continuous 
symmetry leads to the existence of a Goldstone mode with vanishing frequency 
and excitation energy.

The $E_u$ soft mode is Raman inactive in the room temperature centrosymmetric 
pyrochlore structure. Below 200 K, it transforms into $A_1$ and $B_1$ modes, both of which 
can be observed in Raman scattering. The integrated Raman intensities of these 
modes are given by
$I_{i} \propto \langle Q_2 (T) \rangle^2/\omega^2_i(T)$, where $Q_2$ is the 
non-vanishing component of the order parameter at equilibrium and 
$\omega_i$ is the mode frequency~\cite{Fleury72,Cummins83}. The $A_1$ mode 
represents the ordinary \emph{massive} oscillations of the amplitude of the 
order parameter with frequency $\omega_{A_1}(T) \propto Q_2 (T)$ leading to a 
finite value of $I_{A_1}$. Conversely, the Goldstone $B_1$ mode ($\omega_{A_1}(T)=0$) 
should give a zero-frequency peak of divergent intensity. The experimental results 
presented below are in good agreement with this preliminary consideration.


Polarized Raman scattering was measured on surfaces cut and polished along crystal axes. The identical crystal was measured using resonance ultrasound spectroscopy~\cite{Sergienko04}, and Raman results were confirmed on a second crystal. Smooth surfaces reduced the scattered light and allowed for intrinsic intensity measurements very close to the laser line. Consistency between 514.5~nm and 676.4~nm laser excitations confirm the Raman nature of the observed peaks. The experimental resolution was 1~cm$^{-1}$ (676.4~nm) and 3~cm$^{-1}$ (514.5~nm). The focal area was 50$\mu$m~X~500$\mu$m, and the laser power was kept low (40~W/cm$^{2}$) to reduce local heating. Collected photons were dispersed using a Dilor XY500 triple grating spectrometer in subtractive mode (to reduce the stray light) and collected using a LN-cooled CCD. 

By selecting the polarization of incident and scattered photons along certain crystallographic directions, phonons can be identified by symmetry. For the cubic pyrochlore at room temperature, XX polarization (where the first letter indicates the incident and the second the collected vector) measures the $A_{1g} + E_{g}$ channels. The corresponding components of the polarizability tensor $\alpha_{ik}$ are $f_0=(\alpha_{xx}+\alpha_{yy}+\alpha_{zz})/\sqrt{3}$ for $A_{1g}$ and $f_1=(2\alpha_{zz}-\alpha_{xx}-\alpha_{yy})/\sqrt{6}$ and $f_2=(\alpha_{xx}-\alpha_{yy})/\sqrt{2}$ for $E_g$. 
XY polarization (incident and collected photons polarized along orthogonal crystal axes) selects the $F_{2g}$ symmetry ($\alpha_{xy}$, $\alpha_{yz}$, $\alpha_{xz}$). X$^{\prime}$Y$^{\prime}$ polarization (polarized along orthogonal axes 45$^{o}$ rotated from the crystal axes) selects the $E_{g}$ symmetry. Therefore, $F_{2g}$ and $E_{g}$ species are accessed directly while $A_{1g}$ is derived from XX-X$^{\prime}$Y$^{\prime}$. In the tetragonal phases, XX selects $A_{1} + B_{1}$ channels, XY selects $B_{2}$ and X$^{\prime}$Y$^{\prime}$ measures the $B_{1}$ symmetry. $A_{1}$ is derived from XX-X$^{\prime}$Y$^{\prime}$ for this case. 

Below 200 K, tetragonal domains form within the crystal, thereby mixing the X, Y and Z axes. Although no reliable data on the domain size have been reported thus far, the domains are estimated to be of the order of 0.1-10 $\mu$m~\cite{Dodge}, which is smaller than the focal area used in this work.
The presence of these tetragonal domains therefore mixes YY and ZZ polarizations into the nominally $A_1$ and $B_1$ spectra. This mixing, however, does not affect $f_0$, which is invariant under the permutations of $X$, $Y$ and $Z$. Therefore, it is expected that the data derived from XX-X$^{\prime}$Y$^{\prime}$ only corresponds to $f_0$, whereas the X$^{\prime}$Y$^{\prime}$ data contains both $A_1$ and $B_1$ below 200 K. 
For the same reason, ZX and ZY polarizations corresponding to Raman scattering by phonons of $E$ symmetry are seen in the nominally $B_2$ data (not reported here). The selection rules are also compromised by possible slight deviations of the polished planes from the crystal axes. However, strong symmetry dependence is still observed and we can identify mode species by observing the polarization in which the intensity dominates.

\begin{figure}
\vskip-1.2in
\includegraphics[width = 3.5 in]{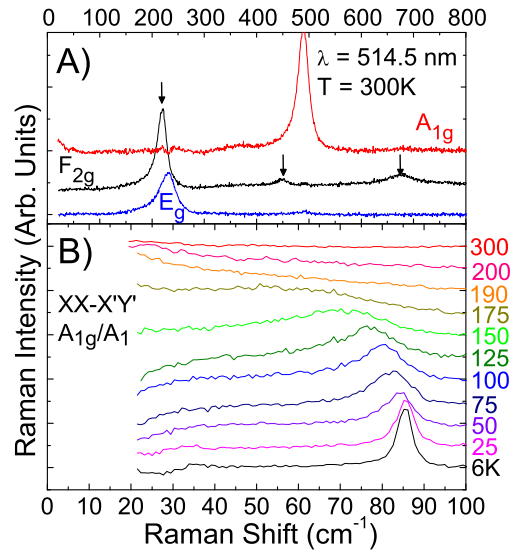}
\vskip-0.7in
\caption{\label{fig2}
(Color online)
(A) Cubic phase (300 K) Raman spectra in $A_{1g}$ (top curve), $F_{2g}$ (middle curve) and $E_{g}$ (lower curve) symmetries. The strongest $F_{2g}$modes are denoted by arrows. (B) XX-X$^{\prime}$Y$^{\prime}$: A$_{1g}$/A$_{1}$ low frequency temperature dependence. The curves have been vertically offset.}
\end{figure}

Raman spectra were measured between our lowest Rayleigh-free frequency (6 cm$^{-1}$ for crossed polarizations XY and X$^{\prime}$Y$^{\prime}$ - 20 cm$^{-1}$ for parallel polarizations XX and X$^{\prime}$X$^{\prime}$) and 800 cm$^{-1}$ 
for the temperatures from 300 K down to 5 K. The symmetry analysis of the Raman active modes in the pyrochlore structure gives one $A_{1g}$, one $E_g$ and four $F_{2g}$ modes~\cite{Sergienko03}. This is in good agreement with the room temperature data presented in Fig. 2(A). Only one of the $F_{2g}$ modes is either too weak or almost degenerate and cannot be resolved.
Below 200 K and 120 K, a large number of new Raman peaks are observed which is consistent with the symmetry lowering 
from centrosymmetric cubic to non-centrosymmetric tetragonal. A detailed report of the high-frequency region of the spectra
will be the subject of a subsequent publication.
For the remainder of this manuscript we concentrate on the symmetry dependent Raman spectra measured at frequencies below 100~cm$^{-1}$. 

In Fig. 2(B) the $A_{1g}/A_1$
low frequency temperature dependence clearly shows the development of the soft mode (the $A_1$ component of $E_u$). It remains overdamped in a small temperature interval below 200 K but 
clearly becomes stiffer as the temperature is lowered to 150 K. We therefore associate this soft mode with the cubic-tetragonal SPT. 
We note that ``parallel'' experimental polarizations, such as XX, generally have strong diffuse scattering which can lead to extrinsic low frequency intensity. Because we do not observe this down to very low frequencies and because such scattering is temperature independent, we conclude that the sample is sufficiently smooth to allow for the resolution of the Goldstone mode. 

\begin{figure}
\includegraphics[width = 3.5 in]{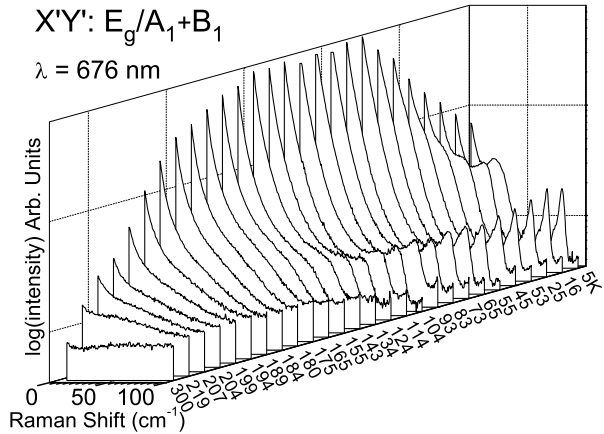}
\caption{\label{fig3}
X$^{\prime}$Y$^{\prime}$: $E_{g}$/$A_1+B_{1}$ symmetry temperature dependence. The log$_{10}$ of the intensity is plotted in 3 dimensional perspective.}
\end{figure}

The log$_{10}$ of the low frequency temperature dependent X$^{\prime}$Y$^{\prime}$ ($E_g$/$A_1+B_1$) Raman intensity is presented 3 dimensionally in Fig. 3. 
Consistent with the presence of different tetragonal domains, the 85 cm$^{-1}$ soft mode observed in $A_{1}$ is clearly evident 
in the X$^\prime$Y$^\prime$ polarization data below 200 K. It first appears as a shoulder of the high-intensity central peak and 
becomes clearly resolved below 155 K. Before turning our attention to the pronounced central peak, we note that another mode (30 cm$^{-1}$ at 6 K) of relatively high
intensity is seen at low temperature. It is a temptation to regard this mode as the soft mode for the 120 K SPT. However,
this fails due to the following arguments. First, it is well known that the 120 K SPT is first-order and hence the soft 
mode behavior is not expected. Second, a soft mode is always expected to generate fully symmetric $A_1$ fluctuations
of the magnitude of the order parameter below the corresponding SPT. In contrast, the 30 cm$^{-1}$ mode practically disappears in
the $A_1$ component obtained by the XX-X$^\prime$Y$^\prime$ subtraction (Fig. 2). 
Therefore, we conclude that it belongs to a low frequency phonon of $B_1$ symmetry which becomes Raman active below 120 K due 
to the symmetry change. 
We cannot simply derive the true temperature dependence of its frequency from our data because a large part of the intensity
is hidden under the central peak. Another obstacle is a complicated fitting procedure that must be employed in the low frequency
region in which the damping constant is comparable to the frequency and, consequently, the line shape is not Lorentzian.

The $B_1$ Goldstone mode is manifested by a large central peak which appears below 
the phase transition at 200 K. Actually, it starts developing slightly above this temperature, which we attribute to the enhanced fluctuations of the order parameter~\cite{Jin02}. The central peak is absent for all temperatures in the $A_{1}$ and $B_{2}$ spectra (to be reported elsewhere). It is also not seen in the room temperature $E_g$ data. Therefore, the usual Rayleigh scattering can be ruled out as its possible origin.
We further conclude that \cd\ does not demonstrate the relaxor behavior which leads to the notorious fully symmetric central peak of finite intensity in certain ferroelectrics~\cite{Jona} due to fluctuations of the same symmetry as the soft mode~\cite{Cowley}. Such a peak would be observable in A$_{1}$ symmetry.

\begin{figure}
\includegraphics[width = 3.5 in]{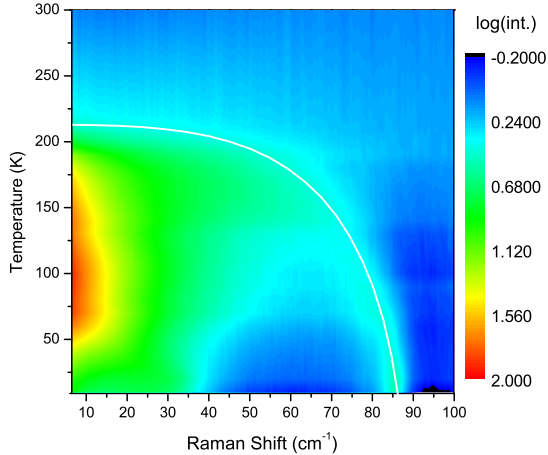}
\caption{\label{}
(Color)
Contour Plot: $E_{g}$/$A_1+B_{1}$ temperature dependence. Log(Intensity) is plotted on a false color scale.}
\end{figure}

Fig. 4 is a contour plot of the $E_{g}$/$A_1+B_{1}$ low frequency temperature dependence. Log(Intensity) is plotted on a false color scale to portray the scattering intensity over a wide dynamic range. The white curve is a guide to the eye highlighting the behavior of the $A_1$ soft mode.  Roughly speaking, the Goldstone mode comprises the red and yellow areas of the Figure. As noted, it begins just above 200K at the lowest frequency and reaches its maximum intensity (height in Fig 3 or width in Fig. 4) at ($\sim$ 85K, well below the lower SPT at 120K. At the lowest temperatures, the intensity goes back down although it remains strong at 6K. 
There are two reasons for the low temperature decrease. First, Raman lines in general tend to narrow at low temperatures because the 
nonlinear mode-mode coupling effects causing damping become less pronounced, and therefore more intensity is concentrated 
around zero frequency. Second, well below the SPT at 200 K, as the order parameter grows in value, one has to take into account
higher order terms in $H_\text{eff}$, which introduce anisotropy and break the continuous symmetry. Under such circumstances, the
mode acquires finite frequency resulting in lower Raman intensity, even though the mode may still remain overdamped.

Further insight into the nature of the Goldstone mode dynamics can be achieved by neutron scattering experiments which 
could fully resolve the central peak at very low frequency and measure the phonon frequency dispersion at finite wave vector 
$\mb k$.
Here we present a simple model which predicts some of the results of neutron diffraction 
in \cd. We consider a three dimensional network of anharmonic oscillators described by the
Hamiltonian~(\ref{Heff}) connected by harmonic springs of stiffness $\kappa$. Every 
oscillator represents a unit cell with the lattice constants $a_\gamma$, $\gamma=x,y,z$.
We introduce the polar coordinates $Q_{1i}=q_i\cos \phi_i$ and $Q_{2i}=q_i\sin \phi_i$, 
where $i$ is the unit cell number. 
Assuming $\alpha < 0$ and $\beta >0$, we approximate the potential energy $\frac\alpha 2 q_i^2 + \frac \beta 4 q_i^4$ by a 
parabola in the vicinity of its minimum $q_0=\sqrt{-\alpha/\beta}$ (see Fig.~\ref{atoms}). The resulting
Hamiltonian is
\begin{eqnarray}
H&=&\sum_i \left[\frac{p_i^2}{2M}+\frac{\pi_i^2}{2Mq_i^2} 
+\frac{M\omega_{A_1}^2}{2}(q_i-q_0)^2 \right] \nn\\
&& + \frac \kappa 2 \sum_{\langle ij \rangle} 
[q_i^2+q_j^2-2q_iq_j\cos(\phi_i-\phi_j)],
\end{eqnarray}
where $p_i$ and $\pi_i$ are the momenta conjugated to $q_i$ and $\phi_i$, respectively, and $\omega_{A_1} = \sqrt{-2\alpha/M}$
is the frequency of the $A_1$ mode at $\mb k = 0$.

The equations of motion can be easily linearized for small deviations from 
the equilibrium value $\phi_i=\phi_j$. 
As a result, we obtain the two branches of dispersion
\begin{eqnarray}
\omega_A(\mb k)&=&(\omega_{A_1}^2+ \frac{4\kappa}{M}\sum_{\gamma}\sin^2 
k_\gamma a_\gamma)^{1/2}\nn\\
\omega_P(\mb k)&=& 2 \sqrt{\frac{\kappa}{M}}(\sum_{\gamma}\sin^2 
k_\gamma a_\gamma)^{1/2}
\end{eqnarray}
for the ``amplitude'' (soft) mode and ``phase'' (Goldstone) mode, respectively. 
For small $\mb k$, $\omega_P(\mb k)$ is linear for any direction in the reciprocal space.
The observation of a mode with linear dispersion in a neutron scattering 
experiment would constitute a valuable validation of the results of Raman scattering 
presented in this Letter.

In summary, we present the evidence of a Goldstone phonon mode in \cd\ obtained by Raman scattering in a wide
temperature range. Using theoretical models, we show that this observation is consistent with the previous experiments
which established the symmetry of the structural order parameter. We also predict the dispersion of the frequency of
the Goldstone mode which can be measured by neutron diffraction.

\begin{acknowledgments}
$^{*}$ These authors contributed equally to this work. 
We thank P. Blaha, S. H. Curnoe, M. D. Lumsden, and D.~J.~Singh for useful discussions.
C. A. K. acknowledges support through ONR/NRL.
I. S. was supported in part by NSERC Canada. 
Oak Ridge National Laboratory is managed by UT-Battelle, LLC, for the U.S. 
Department of Energy under Contract No. DE-AC05-00OR22725.
Work at UT was supported by NSF DMR-007 2998.
\end{acknowledgments}


\begin{thebibliography}{00}

\bibitem{Goldstone62} J. Goldstone \emph{et al.}, \Journal{Phys. Rev.}{127}
{965}{1962}.

\bibitem{Ryder96} L.H. Ryder, \emph{Quantum Field Theory}, 2nd. ed., (Cambridge
Univ. Press, Cambridge, 1996).

\bibitem{Ruegg04} P. B\"oni \emph{et al.}, \Journal{Phys. Rev. B}{52}{10142}
{1995};
I.B. Spielman \emph{et al.}, \Journal{Phys. Rev. Lett.}{87}
{036803}{2001}; Ch. R\"uegg \emph{et al.}, \Journal{Phys. Rev. Lett.}{93}
{257201}{2004}.

\bibitem{Chandra92} I. Mu\v{s}evic, R. Blinc, and B. Zek\v{s},
\emph{The Physics of Ferroelectric and Antiferroelectric Liquid Crystals}
(World Scientific, Singapore, 2000).

\bibitem{Acoustic} Sometimes, the acoustic phonons are regarded as Goldstone 
modes resulting from the breaking of translational symmetry at a hypothetical 
liquid-crystalline solid phase transition. 
Here we deal with a Raman active \emph{optical} Goldstone phonon, which develops
due to a solid-solid structural phase transition.

\bibitem{Hanawa01} M. Hanawa \emph{et al.}, \Journal{Phys. Rev. Lett.}{87}
{187001}{2001}.

\bibitem{Sakai01} H. Sakai \emph{et al.}, \Journal{J. Phys. Condens. Matter}
{13}{L785}{2001}.

\bibitem{Jin01} R. Jin \emph{et al.}, \Journal{Phys. Rev. B}{64}{180503(R)}
{2001}.

\bibitem{Castellan02} J.P. Castellan \emph{et al.}, \Journal{Phys. Rev. B}
{66}{134528}{2002}.

\bibitem{Yamaura02} J.-I. Yamaura and Z. Hiroi, \Journal{J. Phys. Soc. Jpn.}
{71}{2598}{2002}.

\bibitem{Arai02} K. Arai \emph{et al.}, \Journal{J. Phys. Condens. Matter}{14}
{L461}{2002}. 

\bibitem{Sergienko03} I.A. Sergienko and S.H. Curnoe, \Journal{J. Phys. Soc. 
Jpn.}{72}{1607}{2003}.


\bibitem{Sergienko04} I.A. Sergienko \emph{et al.}, \Journal{Phys. Rev. Lett.}
{92}{065501}{2004}.

\bibitem{Lu05} C. Lu \emph{et al.}, \Journal{Phys. Rev. B}{70}{092506}{2004}.

\bibitem{Fleury72} P.A. Fleury, \Journal{Comments Solid State Phys.}{4}{167}
{1972}.

\bibitem{Cummins83} V.L. Ginzburg \emph{et al.} in \emph{Light Scattering near 
Phase Transitions}, Eds. H.Z. Cummins and A.P. Levanyuk (North-Holland, 
Amsterdam, 1983), p. 3.


\bibitem{Dodge} J. S. Dodge and B. D. Gaulin, private communication.

\bibitem{Jona} O. Svitelskiy \emph{et al.}, \Journal{Phys. Rev. B}{68}{104107}{2003}.

\bibitem{Cowley} A. D. Bruce and R. A. Cowley, \Journal{Adv. Phys.}{29}{219}
{1980}.

\bibitem{Jin02} R. Jin \emph{et al.}, \Journal{J. Phys.: Condens. Matter}
{14}{L117}{2002}.

\end{thebibliography}
\end{document}